\documentclass[prd,twocolumn,showpacs,floats,letterpaper,nofootinbib]{revtex4}
\usepackage{amssymb,amsmath,amsfonts}
\usepackage{color}
\usepackage{graphicx}

\newcommand{\wthrj}[6]{\left(
                           \begin{array}{ccc}
        \! #1\! & #2\!  & #3\!  \\
        \! #4\! & #5\!  & #6\!
                           \end{array}
			   \right)}
\newcommand{\wsixj}[6]{\left\{
                           \begin{array}{ccc}
         #1 & #2  & #3  \\
         #4 & #5  & #6
                           \end{array}
			   \right\}}

\newcommand{\intl}[1]{\int {d^2 {\bf l}_#1 \over (2\pi)^2}}

\newcommand{\intlp}{\int {d^2 {\bf l'}\over (2\pi)^2}}
\newcommand{\cmb}{{\Theta}}

\newcommand{\len}{\phi}

\newcommand{\bea}{\begin{eqnarray}}
\newcommand{\eea}{\end{eqnarray}}
\newcommand{\bean}{\begin{eqnarray*}}
\newcommand{\eean}{\end{eqnarray*}}

\newcommand{\bx}{{\bf x}}

\newcommand{\rad}{{r}}
\newcommand{\da}{{d_A}}
\newcommand{\bk}{{\bf k}}
\newcommand{\bn}{{\bf \hat{n}}}

\newcommand{\vecla}{{{\bf l}_1}}

\newcommand{\nn}{\nonumber\\}
\newcommand{\bfl}{{\mathbf{l}}}

\newcommand{\fnl}{{f_{\rm NL}}}
\newcommand{\lp}{l^{\prime}}
\newcommand{\ldp}{l^{\prime \prime}}
\newcommand{\ltp}{l^{\prime \prime \prime}}

\newcommand{\ltwop}{l_2^{\prime}}
\newcommand{\ltwodp}{l_2^{\prime \prime}}

\newcommand{\lthrp}{l_3^{\prime}}
\newcommand{\lthrdp}{l_3^{\prime \prime}}
\newcommand{\lthrtp}{l_3^{\prime \prime \prime}}
\newcommand{\mpr}{m^{\prime}}
\newcommand{\mdp}{m^{\prime \prime}}
\newcommand{\mtp}{m^{\prime \prime \prime}}

\newcommand{\mtwopr}{m_2^{\prime}}
\newcommand{\mtwodp}{m_2^{\prime \prime}}

\newcommand{\mthrpr}{m_3^{\prime}}
\newcommand{\mthrdp}{m_3^{\prime \prime}}
\newcommand{\mthrtp}{m_3^{\prime \prime \prime}}
\newcommand{\la}{\langle}
\newcommand{\ra}{\rangle}

\newcommand{\ApJ}{Astrophys. J.}
\newcommand{\PRL}{Phys. Rev. Lett.}
\newcommand{\PRD}{Phys. Rev. D}

\begin{document}

\title{Weak Lensing of  the Primary CMB Bispectrum} 
\author{Asantha Cooray, Devdeep Sarkar, and  Paolo Serra} 
\affiliation{Center for Cosmology,
Department of Physics  and  Astronomy, 
University  of California, Irvine, CA 92697}

\begin{abstract}
The cosmic  microwave background (CMB) bispectrum is a
well-known probe  of the non-Gaussianity  of primordial perturbations.
Just as the intervening large-scale structure modifies the CMB angular
power spectrum through weak  gravitational lensing, the CMB primary
bispectrum generated  at the last scattering surface  is also modified
by lensing. We discuss the  lensing modification to the CMB bispectrum
and show that lensing leads to an overall decrease in the amplitude of
the primary  bispectrum at multipoles  of interest between  100 and
2000 through  additional smoothing introduced by  lensing.  Since weak
lensing  is  not accounted for in  current  estimators of  the
primordial  non-Gaussianity parameter,  the  existing measurements  of
$f_{\rm NL}$  of the local  model with WMAP  out to $l_{\rm  max} \sim
750$ is  biased low  by about 6\%.   For a high  resolution experiment
such as  Planck, the  lensing modification to  the bispectrum  must be
properly  included when  attempting to  estimate the  primordial
non-Gaussianity or the bias will be  at the level of 30\%. For Planck,
weak  lensing   increases  the   minimum  detectable  value   for  the
non-Gaussianity parameter of the local type $f_{\rm NL}$ to 7 from the
previous estimate of about  5 without lensing.  The minimum detectable
value of $f_{\rm NL}$ for a cosmic variance limited experiment is also
increased from less than 3 to $\sim$ 5.
\end{abstract}

\pacs{98.70.Vc,98.65.Dx,95.85.Sz,98.80.Cq,98.80.Es}

\maketitle

\section{Introduction} 
The  weak  lensing of  cosmic  microwave  background (CMB)  anisotropy
angular  power  spectrum is  now  well  understood  in the  literature
\cite{lensing,Lewis}.  The modifications result  in a smoothing of the
acoustic peak  structure at  large angular scales  and an  increase in
power below  a few arcminute  angular scales corresponding to  the the
damping   tail  of   CMB   anisotropies  \cite{Hu00}.    

The angular  power spectrum  of the lensing potential out to
the last  scattering surface can be 
established with   quadratic estimators that probe lensing non-Gaussianity
at the 4-point level of a CMB map \cite{HuOka02}.  Such a reconstruction of the lensing
potential is helpful  for CMB B-mode  studies of
polarization   \cite{KamKosSte97},  especially   in  the   context  of
searching   for    the   signature   of    the primordial   tensor   modes
\cite{KesCooKam02}.   This is  due to  the  fact that  in addition  to
inflationary gravitational waves, the B-modes of CMB polarization also
contains a signal  generated by lensing of scalar  E-modes with a peak
in     power     at     a     few     arcminute     angular     scales
\cite{Zaldarriaga:1998ar}. The lensing reconstruction has now been applied
to the existing  Wilkinson Microwave Anisotropy Probe  (WMAP) data leading
to a $\sim$ 2$\sigma$  to 3$\sigma$ detection of gravitational lensing
in  the CMB through  a correlation  between the  reconstructed lensing
potential  and tracers  of  the large-scale  structure  such as  radio
galaxies \cite{Smith,Hirata}.

In parallel  with the  progress on lensing  studies with the  CMB, the
search for  primordial non-Gaussianity  using the CMB  bispectrum with
constraints on  the non-Gaussianity parameter  $f_{\rm NL}$ is  now an
active  topic in  cosmology  \cite{Komatsu,Lig,Komatsu4}.  The  5-year
WMAP data is consistent with $-9  < f_{\rm NL} <111$ at the 95\%
confidence  level  for the  local  model  \cite{Komatsu5yr}, though  a
non-zero  detection of  primordial  non-Gaussianity at  the same  95\%
confidence level with $26.9 < f_{\rm NL} < 146.7$ is claimed elsewhere
using the WMAP 3-year data \cite{Yadav}.  This result, if correct, has
significant cosmological  implications since the  expected value under
standard   inflationary   models    is   $f_{\rm   NL}   \lesssim   1$
\cite{Salopek,Falk,Gangui,Pyne,Acquaviva,Maldacena,Bartolo},     though
alternative  models  of inflation,  such  as  the ekpyrotic  cosmology
\cite{Buchbinder,Lehners},   generally  predict  a   large  primordial
non-Gaussianity with $f_{\rm NL}$ at few tens.

Just as the  CMB power spectrum is modified  by lensing from potential
fluctuations of the  intervening large-scale structure \cite{lensing},
the  CMB bispectrum will  also be  modified by  gravitational lensing.
The  correlation  between  the  projected lensing  potential  and  CMB
secondary effects, such as  the integrated Sachs-Wolfe (ISW) effect or
the Sunyaev-Zel'dovich (SZ) effect,  leads to a non-Gaussian signal at
the   three  point   level  \cite{Goldberg,Cooray}.   These  secondary
non-Gaussianities  are expected even  if the  primordial perturbations
are Gaussian and  impact existing primordial non-Gaussianity parameter
measurements   by   introducing  a   small,   but  unavoidable,   bias
\cite{Smith2,Serra,BabichPier}.

Beyond  secondary non-Gaussianities, weak  lensing by  the intervening
large-scale structure maps the intrinsically non-Gaussian CMB sky to a different
anisotropy  pattern  when observed  today.   Thus  the bispectrum  one
reconstructs  with a  CMB map,  assuming it  to be  of the
expected  form  at  the  last-scattering  surface  due  to  primordial
non-Gaussian perturbations, will result  in a biased estimate of
the primordial non-Gaussianity  parameter.  The existing estimator can
be  modified to  account for  lensing  modifications and  to obtain  a
bias-free  estimate of  the  non-Gaussianity, but  at  the expense  of
factorizability that has allowed fast computation of the bispectrum in
existing  analyses   \cite{Komatsu2}.   Since  lensing   modifies  the
anisotropy  pattern by  smoothing the  fluctuations, a  change  in the
minimum detectable non-Gaussianity parameter  $f_{\rm NL}$ for a given
experiment is expected to be different from the existing values in the
literature \cite{Komatsu}.

In  this paper,  we present  a general  derivation of  the  lensed CMB
primary  bispectrum  and  quantify  above statements  on  the  changes
imposed by  lensing for  the detection of  primordial non-Gaussianity.
We find  that the non-Gaussianity parameter  measured with WMAP,
ignoring lensing, will result in an estimate of $f_{\rm NL}$ for
the local model that is biased low by about 6\%, when measurements are
extended out  to $l_{\rm max}  \sim 750$.  Furthermore,  with lensing,
the minimum detectable level of  $f_{\rm NL}$ with Planck is increased
by roughly  40\% from less than 5  to about 7 and  the cosmic variance
limit of $f_{\rm NL}$ is increased from 3 to 5.

This  paper  is  organized  as  follows:  we  first  discuss  the  CMB
primary  bispectrum of the  local type  in Section~II.   Some basic
ingredients  related  to  the  lensing  calculation  is  presented  in
Section~III.  We derive the  lensing effect  on the  bispectrum, under
both flat-sky and all-sky formulations, in Section~IV.  We discuss our
results and conclude with a  summary in Section~V. In illustrating our
results  we make use  of the  standard flat  $\Lambda$CDM cosmological
model  consistent with  WMAP with  $\Omega_b=0.042$, $\Omega_c=0.238$,
$h=0.732$, $n_s=0.958$ and $\tau=0.089$.

\section{ CMB Primary Bispectrum}

The  CMB  temperature   perturbation  on  the  sky,  $\cmb(\bn)=\Delta
T(\bn)/T$, is decomposed into its multipole moments 
\begin{equation}
\cmb(\bn) = \sum_{lm} \cmb_{lm} Y_{l}^{m}(\bn) \, .
\end{equation}
The  angular power  spectrum and  bispectrum of  CMB  anisotropies are
defined in the usual way, respectively, as
\begin{eqnarray}
\langle \cmb_{l m} \cmb_{l' m'} \rangle &=& \delta_{l,l'}\delta_{m,m'} C_{l}^{\cmb \cmb} \,, \nn
\langle \cmb_{l_1m_1} \cmb_{l_2m_2} \cmb_{l_3m_3} \rangle &=& \wthrj{l_1}{l_2}{l_3}{m_1}{m_2}{m_3} B_{l_1 l_2 l_3}\,,
 \label{eqn:pscmb}
\end{eqnarray}
where, for the bispectrum, we have  introduced the Wigner-3$j$ symbol (see  the Appendix of
Ref.~\cite{Cooray}    for    some    useful   properties    of    this
symbol).  

The CMB bispectrum is generated by a coupling of the local-type with a
quadratic correction to the Newtonian curvature such that
\begin{equation}
\Phi(\bx)=\Phi_L(\bx)+\fnl \left[\Phi^2_L(\bx)-\langle \Phi^2_L(\bx)\rangle\right] \, 
\label{phi}
\end{equation}
where $\Phi_L(\bx)$ is the  linear and Gaussian perturbation and
$\fnl$ in  the non-Gaussianity parameter,  which is taken to  be scale
independent \cite{Komatsu}.

In Fourier space, we can decompose equation~(\ref{phi}) as
\begin{equation}
\Phi(\bk)=\Phi_L(\bk)+f_{\rm NL} \Phi_{\rm NL}(\bk) \, ,
\end{equation}
with
\begin{eqnarray}
\Phi_{\rm NL}(\bk) &=& \int \frac{d^3{\bk_1}}{(2\pi)^3}\Phi_L(\bk+\bk_1) \Phi_L^*(\bk_1) \nn
 && - (2\pi)^3 \delta(\bk)\int \frac{d^3{\bk_1}}{(2\pi)^3}P_{\Phi}(k_1)\, ,
\end{eqnarray}
where $P_{\Phi}(k)$ is the linear power spectrum, defined as
\begin{equation}
\langle \Phi(\bk) \Phi(\bk') \rangle = (2\pi)^3 \delta(\bk+\bk')P_{\Phi}(k) \, .
\end{equation}

The  multipole moments  of  the  anisotropy  can be written as
\begin{equation}
  \label{eq:almphi}
  \cmb_{lm}=4\pi(-i)^l \int\frac{d^3{\mathbf k}}{(2\pi)^3}\Phi({\mathbf k})g_{Tl}(k)Y_{lm}^*(\hat{\mathbf k}),
\end{equation}
where  $\Phi({\mathbf  k})$ from  above  is  the primordial  curvature
perturbation in  the Fourier space,  and $g_{Tl}(k)$ is  the radiation
transfer  function.   With  $\Phi_L(\bk)$  and  $\Phi_{\rm  NL}$,  the
moments   can   be    separated   into   two   components   with
$\cmb_{lm}=\cmb_{lm}^L + \cmb_{lm}^{NL}$.

The  CMB angular  power spectrum  can  be defined  using the  transfer
function and the power spectrum of dominant linear fluctuations as
\begin{equation}
C_l^\cmb = \frac{2}{\pi} \int_0^\infty k^2 dk P_{\Phi}(k)g^2_{Tl}(k) \, .
\end{equation}

Using     the     definition     of     the     angular     bispectrum
(equation~\ref{eqn:pscmb}),  the   primordial  temperature  anisotropy
bispectrum can be written as
\begin{eqnarray}
  \nonumber
&&  B_{l_1l_2l_3}
  =  \sum_{m_1m_2m_3} 
\left(
  \begin{array}{ccc}
  l_1&l_2&l_3\\
  m_1&m_2&m_3
  \end{array}
  \right)\bigg[
  \langle \cmb_{l_1m_1}^L\cmb_{l_2m_2}^L\cmb_{l_3m_3}^{NL}\rangle \nn
  && + \langle \cmb_{l_1m_1}^L\cmb_{l_2m_2}^{NL}\cmb_{l_3m_3}^{L}\rangle + \langle \cmb_{l_1m_1}^{NL}\cmb_{l_2m_2}^L\cmb_{l_3m_3}^{L}\rangle \bigg] \, ,
\end{eqnarray}
which can be simplified to \cite{Komatsu}
\begin{eqnarray}
B_{l_1l_2l_3}  &=& 2{\cal G}_{l_1l_2l_3}\int_0^\infty r^2 dr b^L_{l_1}(r)b^L_{l_2}(r)b^{NL}_{l_3}(r)  \label{eq:almspec} \\ 
&& + b^L_{l_1}(r)b^{NL}_{l_2}(r)b^{L}_{l_3}(r)+ b^{NL}_{l_1}(r)b^L_{l_2}(r)b^{L}_{l_3}(r) , \nonumber
\end{eqnarray}
where 
\begin{eqnarray}
  \label{eq:bLr}
  b^L_{l}(r) &=&
  \frac2{\pi}\int_0^\infty k^2 dk P_\Phi(k)g_{Tl}(k)j_l(kr),\\
  \label{eq:bNLr}
  b^{NL}_{l}(r) &=&
\fnl \frac2{\pi}\int_0^\infty k^2 dk g_{Tl}(k)j_l(kr) \, ,
\end{eqnarray}
and
\begin{equation}
  \label{eq:wigner*}
  {\cal G}_{l_1l_2l_3}
  = \sqrt{\frac{(2l_1+1)(2l_2+1)(2l_3+1)}{4\pi}}
  \left(
  \begin{array}{ccc}
  l_1&l_2&l_3\\
  0&0&0
  \end{array}
  \right)\, .
\end{equation}

When illustrating our results, we make use of a 
modified code of CMBFAST \cite{Seljak} for the standard flat $\Lambda$CDM cosmological model
to fully calculate radiation transfer functions when generating the  CMB primary bispectrum.

\begin{figure}[!t]
\includegraphics[scale=0.38,angle=-90]{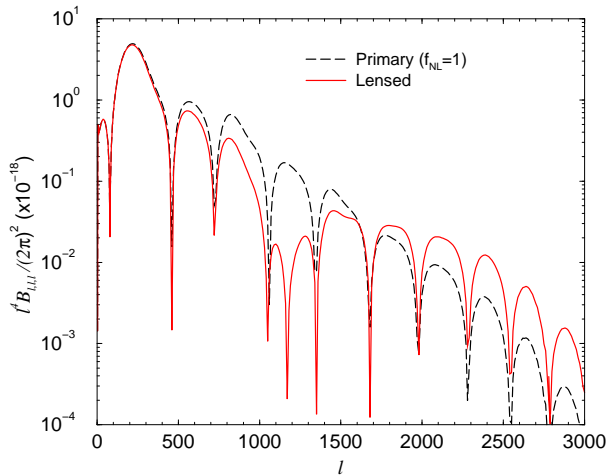}
\caption{The   CMB   bispectrum   for  the   equilateral   case ($l_1=l_2=l_3=l$)   with  (solid  line)   and  without   (dashed  line)
lensing. Here  we plot  $l^4B^\theta_{lll}/(2\pi)^2$ as a  function of
the multipole $l$ for one of  the sides. We assume $f_{\rm NL}=1$. The
lensing effect can be described as  a decrease in the amplitude of the
bispectrum  when  $l  \lesssim   1700$  with  an  increase  at  higher
multipoles.}
\label{cl}
\end{figure}

\section{Weak Lensing Basics}

The  effects of  weak  lensing  can be  encapsulated,  under the  Born
approximation, in the radial projection of the gravitational potential
($\Phi$), given as \cite{cosmicshearrefs}
\begin{eqnarray}
\phi(\bn)&=&- 2 \int_0^{\rad_s} d\rad'
\frac{\da(\rad_s-\rad')}{\da(\rad_s)\da(\rad')} \Phi (\mathbf{r}(\hat{{\bf n}}),\rad')\,,
\label{eqn:lenspotential}
\end{eqnarray}
where $\rad(z)$  is the line-of-sight comoving  distance (or look-back
time)  to  a redshift  $z$  from  the  observer with  last  scattering
surface  at   $\rad_s=\rad(z=1100)$,  and  $\da(\rad)$   is  the
comoving  angular diameter  distance.  In  a spatially  flat universe,
$\da \rightarrow \rad$.   Here, we ignore the time-delay  effect as it
is  small  compared  to  the  geometric  lensing  effect  captured  by
equation~(\ref{eqn:lenspotential}) \cite{CooHutime}.

The  calculation  related to  the  CMB  bispectrum described  below
requires the angular power spectrum of lensing potential $\phi$, which can be
decomposed into the multiple moments as
\begin{equation}
\phi(\bn) = \sum_{lm} \phi_{lm} Y_{l}^{m}(\bn) \, ,
\end{equation}
with the lensing power spectrum defined using
$\langle \phi_{lm} \phi_{\lp \mpr}\rangle = \delta_{l, \lp}\delta_{m, \mpr}C^{\phi}_{l}$
to obtain \cite{Hu00}
\begin{equation}
        C_l^\phi = \frac{2}{\pi} \int k^2 dk P_{\Phi}(k) [I_l^{\rm len}(k)]^2 \,,
\label{eqn:clphi}
\end{equation}
where
\begin{eqnarray}
I_l^{\rm len}(k)& =&
                \int d \rad\, W^{\rm len}(\rad)
                 j_l(k\rad)  \,,\nonumber\\
W^{\rm len}(\rad)& =&
                -2 F(\rad) {\da(\rad_s-\rad) \over
                \da(\rad) \da(\rad_s)}\,.
\label{eqn:lensint}
\end{eqnarray}
Here   $F(\rad)$   describes  the   radial   evolution  of   potential
fluctuations.  Modifications  to the CMB  anisotropies, generated
at higher order in lensing potential fluctuations, are at the level of
at most  5\% relative  to those due  to the lensing  potential angular
power spectrum \cite{Cooray3}.   The bispectrum of lensing potentials,
due  to  the  non-linear  evolution  of  density  perturbations,  also
modifies the  CMB primary  bispectrum, but these  changes can  also be
ignored  since  the  lensing  potential  bispectrum is  at  the  order
$\left(C_l^\phi\right)^2$, while  changes we describe  are first order
in the  angular power  spectrum of the  lensing potential.   Using the
Limber   approximation,  equation~(\ref{eqn:clphi})  can   be  further
simplified, but we do not  make use of this approximation in numerical
calculations illustrated here since the flat-sky form of the potential
power spectrum is known to  bias lensing results of the power spectrum
by about 10\% at all multipoles \cite{Hu00}.

\section{Lensing of the CMB Bispectrum}

We  first  give a  treatment  of the  lensing  of  the CMB  bispectrum
assuming a  flat-sky approximation and discuss a  derivation under the
spherical sky later.

\begin{figure*}[!t]
\includegraphics[scale=0.38,angle=-90]{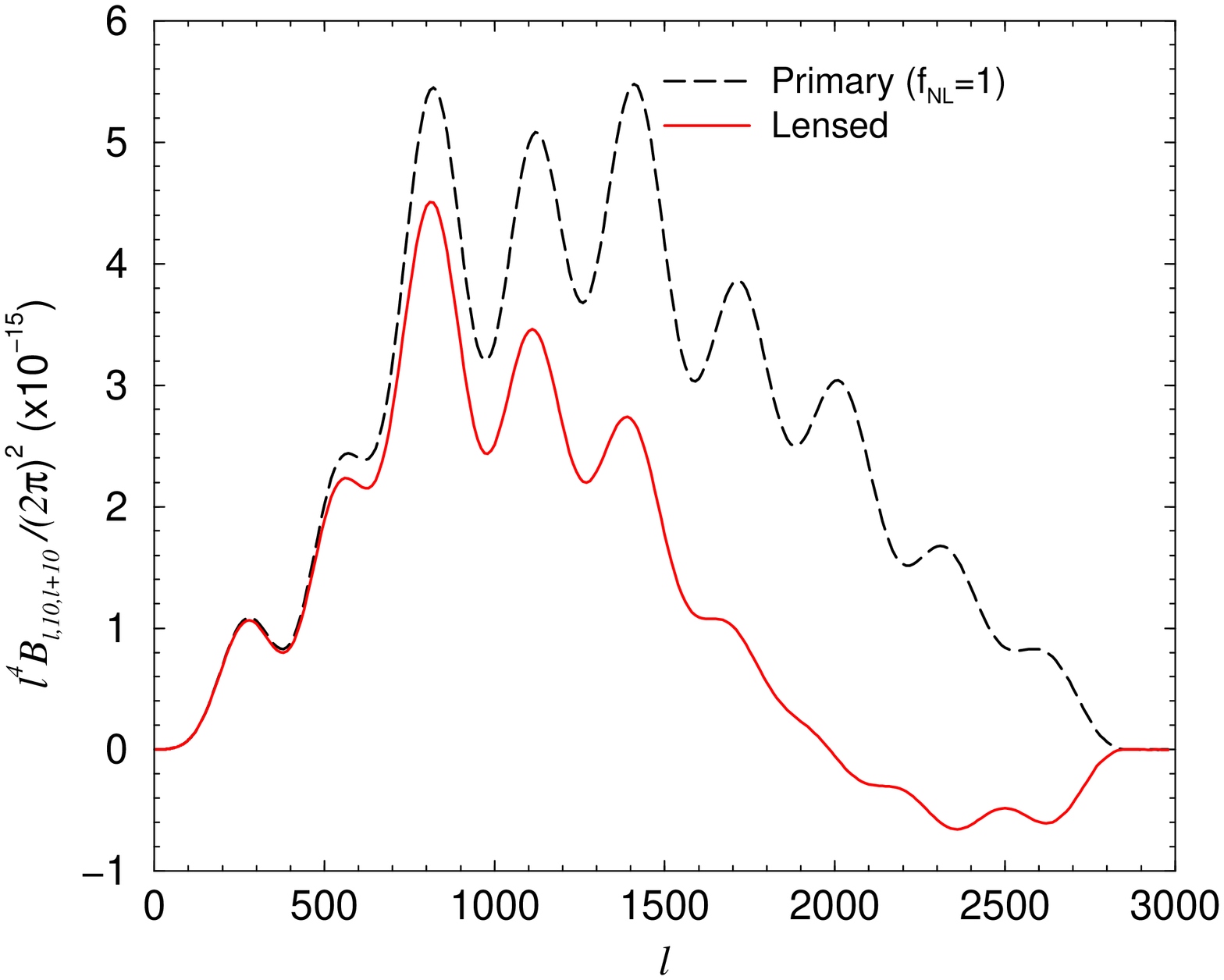}
\includegraphics[scale=0.38,angle=-90]{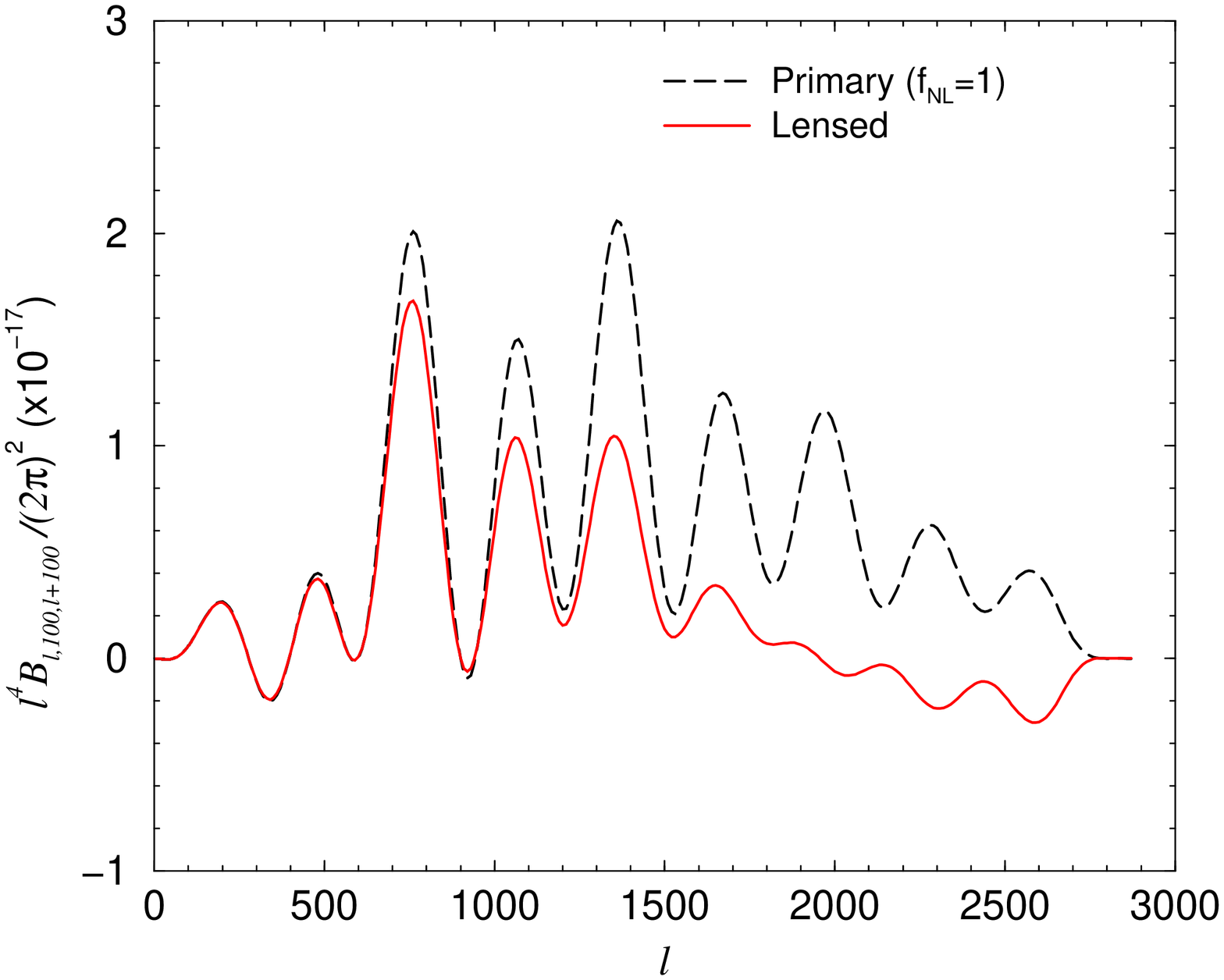}
\caption{The squeezed configurations  ($l_1\sim l_3 \gg l_2$) of
 the  CMB  bispectrum with  (solid  line)  and  without (dashed  line)
 lensing.   The  left  panel  is  for  $l_2=10$  and  right  panel  is
 $l_2=100$. We  vary $l=l_1$ with  $l_3=l+l_2$ in both cases  and plot
 $l^4B_{l,l_2,l+l_2}/(2\pi)^2$ as  a function  of $l$. Again,  we take
 $f_{\rm NL}=1$.   In these configurations, the lensing  effect can be
 described as an  overall decrease in the amplitude  of the bispectrum
 when $l \lesssim 1200$. This suggests that lensing by the intervening
 large-scale  structure leads  to a  less non-Gaussianity  in  the CMB
 map.}
\label{cl2}
\end{figure*}

\subsection{Flat-sky Case}

Weak lensing  deflects the path  of background photons resulting  in a
remapping of the observed anisotropy pattern on the sky.  Following an
approach  similar  to \cite{Hu00},  we  write  the lensed  temperature
anisotropy as
\begin{eqnarray}
\tilde \cmb(\bn)  & = &   \cmb[\bn + \nabla\len(\bn)] \nn
 & \approx & \cmb(\bn) + \nabla_i \len(\bn) \nabla^i \cmb(\bn) \nn
 & & + {1 \over 2} \nabla_i \len(\bn) \nabla_j \len(\bn) \nabla^{i}\nabla^{j} \cmb(\bn) + \ldots  \;
\label{eqn:pert}
\end{eqnarray}
Here, $\cmb(\bn)$ is the  unlensed CMB temperature anisotropy, $\tilde
\cmb(\bn)$  is the  lensed  anisotropy, and  $\nabla\len(\bn)$ is  the
lensing  deflection angle  for the  CMB photons.

Taking the Fourier transform, as  appropriate for a flat-sky, we write
the lensed temperature anisotropy in Fourier space as
\begin{eqnarray}
\tilde \cmb(\vecla)
&=& \int d \bn\, \tilde \cmb(\bn) e^{-i \vecla \cdot \bn} \nonumber \nn
&=& \cmb(\vecla) - \intl{1'} \cmb(\vecla') L(\vecla,\vecla')\,,
\label{eqn:thetal}
\end{eqnarray}
where
\begin{eqnarray}
\label{eqn:lfactor}
L(\vecla,\vecla') &\equiv & \len(\vecla-\vecla') \, (\vecla - \vecla') \cdot \vecla' \nn
& &-{1 \over 2} \intl{1''} \len(\vecla'') \len(\vecla - \vecla' - \vecla'') \, \nn
& & \quad \times (\vecla'' \cdot \vecla') (\vecla - \vecla' - \vecla'')\cdot \vecla'  \, 
\end{eqnarray}
to the second order in lensing potential in the perturbative expansion.

The  observed angular power  spectrum of  CMB anisotropies  under weak
 lensing is  discussed in  \cite{Hu00}.  The resulting  power spectrum
consists of both the unlensed intensity and a perturbative correction
 related to the lensing effect.  Making use of the expansion and after
 some straight  forward calculations, we obtain  the lensed anisotropy
 power spectrum as
\begin{equation}
\tilde C^{\cmb}_{l}=C^{\cmb}_{l} \left(1- l^2 R\right) + \intl{1} C^{\phi}_{l_1} C^{\cmb}_{| \bfl - \bfl_1|} [(\bfl - \bfl_1)\cdot \bfl_1]^2,
\label{eqn:ttflat}
\end{equation}
where 
\begin{equation}
R ={1 \over  4\pi} \int dl\; l^3\; C_{l}^{\phi}\, .
\end{equation}
Here,  $R$  describes  the  variance  of the  deflection  angle.   For
$\Lambda$CDM     cosmology,    $\theta_{\rm    rms}=\sqrt{R}\sim
2.6'$. This derivation makes use of the flat-sky approximation to
describe the lensing effect on CMB anisotropy power spectrum.
When the expressions  derived in  the previous  section for
$C_l^\cmb$ and $C_l^\phi$ under  the exact spherical-sky treatment are used in 
equation~(\ref{eqn:ttflat}),  the lensed CMB power spectrum can be derived with
a less bias than  using, say, the flat-sky  result  for  $C_l^\phi$ in the same
expression \cite{Hu00}.

Keeping the flat-sky approximation, we can define the angular bispectrum as
\begin{equation}
\langle \cmb(\bfl_1) \cmb(\bfl_2) \cmb(\bfl_3)\rangle \equiv (2\pi)^2 \delta(\bfl_1+\bfl_2+\bfl_3) B^{\cmb}_{(\bfl_1,\bfl_2,\bfl_3)}\,,
\end{equation}
and following the approach similar to the lensed angular power spectrum that
led to equation~(\ref{eqn:ttflat}), the lensed bispectrum can be expressed as
\begin{eqnarray}
&&\tilde B^{\cmb}_{(\bfl_1,\bfl_2,\bfl_3)} = \\
&& B^{\cmb}_{(\bfl_1,\bfl_2,\bfl_3)}\left[1-\left(l_1^2+l_2^2+l_3^2\right) \frac{R}{2} \right] +\intlp C^{\phi}_{l'} \nn
&& \quad \times \Big[B^{\cmb}_{(\bfl_1,\bfl_2-\bfl',\bfl_3+\bfl')}(\bfl_2-\bfl')\cdot \bfl'(\bfl_1+\bfl_2-\bfl')\cdot \bfl' \nn
&& \quad \quad + B^{\cmb}_{(\bfl_1-\bfl',\bfl_2+\bfl',\bfl_3)}(\bfl_1-\bfl')\cdot \bfl'(\bfl_3+\bfl_1-\bfl')\cdot \bfl' \nn
&& \quad \quad + B^{\cmb}_{(\bfl_1+\bfl',\bfl_2,\bfl_3-\bfl')}(\bfl_3-\bfl')\cdot \bfl'(\bfl_3+\bfl_2-\bfl')\cdot \bfl'  \Big] \, \nonumber.
\end{eqnarray}
Note   that   we   have   identified  the   flat-sky   bispectrum   as
$B^{\cmb}_{(\bfl_1,\bfl_2,\bfl_3)}$  to distinguish  from  the all-sky
bispectrum  $B^{\cmb}_{l_1,l_2,l_3}$.   The  two are  related
through
\begin{equation}
B^{\cmb}_{l_1 l_2 l_3} = \wthrj{l_1}{l_2}{l_3}{0}{0}{0} \sqrt{\frac{(2l_1+1)(2l_2+1)(2l_3+1)}{4\pi}} B^{\cmb}_{(\bfl_1,\bfl_2,\bfl_3)}
\end{equation}

\subsection{All-sky Treatment}

The derivation related to the  lensing of the CMB bispectrum under the
more  appropriate  spherical sky  can  be  obtained  by replacing  the
Fourier components with spherical harmonic multipole moments.  In this
case, the lensed field can be represented as \cite{Hu00}
\begin{eqnarray}
\tilde \cmb_{l m} &\approx& \cmb_{l m} + \int d\bn Y_l^{m*} \nabla_i \len(\bn) \nabla^i \cmb(\bn) \nonumber\\
&& \quad + {1 \over 2} 
\int d\bn Y_l^{m*} 
\nabla_i \len(\bn) \nabla_j \len(\bn) \nabla^{i}\nabla^{j} \cmb(\bn)
\nonumber\\
& = & \cmb_{l m} + \sum_{\lp \mpr} \sum_{\ldp \mdp} 
			\len_{\lp \mpr}  \cmb_{\ldp \mdp} 
\label{eqn:thetalm}
\\ && \quad \times			
\bigg[ I_{l \lp \ldp}^{m \mpr \mdp} 
+ {1 \over 2}  
\sum_{\ltp \mtp} 
			\len_{\ltp \mtp}^* 
J_{l \lp \ldp \ltp}^{m \mpr \mdp \mtp} \bigg] \,, \nonumber
\end{eqnarray}
where, the  integrals over the  spherical harmonics were  replaced, in
the last step, by the geometrical factors
\begin{eqnarray}
I_{l \lp \ldp}^{m \mpr \mdp} &=&
			\int d\bn \, Y_l^{m*} \left( \nabla_i Y_{\lp}^{\mpr} \right)
					      \left( \nabla^i Y_{\ldp}^{\mdp} \right)\,, \\
\label{eqn:IJform}
J_{l \lp \ldp \ltp}^{m \mpr \mdp \mtp}
			 &=& \int d\bn\, Y_l^{m*}  
				\left( \nabla_i  Y_{\lp}^{\mpr}       \right)
				\left( \nabla_j Y_{\ltp}^{\mtp *} \right)
				 \nabla^{i}\nabla^j Y_{\ldp}^{\mdp}      \,.\nonumber
\end{eqnarray}

Using   equation~(\ref{eqn:pscmb}),   the   lensed   CMB   temperature
bispectrum can then be expressed as
\begin{eqnarray}
\tilde{B}_{l_1 l_2 l_3}^{\cmb} &=& \sum_{m_1 m_2 m_3} \wthrj{l_1}{l_2}{l_3}{m_1}{m_2}{m_3}  \langle \tilde{\cmb}_{l_1 m_1} \tilde{\cmb}_{l_2 m_2} \tilde{\cmb}_{l_3 m_3} \rangle  \, ,\nn
\end{eqnarray}
leading to
\begin{eqnarray}
&& \tilde{B}_{l_1 l_2 l_3}^{\cmb} = \\ 
&&  \sum_{m_1 m_2 m_3}  \wthrj{l_1}{l_2}{l_3}{m_1}{m_2}{m_3}  \bigg[  \langle {\cmb}_{l_1 m_1} {\cmb}_{l_2 m_2} {\cmb}_{l_3 m_3} \rangle \nn
&& + \frac{1}{2} \sum_{\lthrp \mthrpr} \sum_{\lthrdp \mthrdp} \sum_{\lthrtp \mthrtp} \la {\cmb}_{l_1 m_1} {\cmb}_{l_2 m_2} \cmb_{\lthrdp \mthrdp} \nn
&&  \hspace{3mm} \phi_{\lthrp \mthrpr} \phi^{*}_{\lthrtp \mthrtp} \ra J_{l_3 \lthrp \lthrdp \lthrtp}^{m_3 \mthrpr \mthrdp \mthrtp}  +{\rm 2 \; Perm.}\nn
&& + \sum_{\ltwop \mtwopr} \sum_{\ltwodp \mtwodp}  \sum_{\lthrp \mthrpr} \sum_{\lthrdp \mthrdp} \la {\cmb}_{l_1 m_1}  \cmb_{\ltwodp \mtwodp} \cmb_{\lthrdp \mthrdp} \nn
&& \hspace{3mm} \phi_{\ltwop \mtwopr} \phi_{\lthrp \mthrpr} \ra I_{l_2 \ltwop \ltwodp}^{m_2 \mtwopr \mtwodp} I_{l_3 \lthrp \lthrdp}^{m_3 \mthrpr \mthrdp} + {\rm 2 \; Perm.} \bigg] \, . \nonumber
\end{eqnarray}

\begin{figure*}[!t]
\includegraphics[scale=0.38,angle=-90]{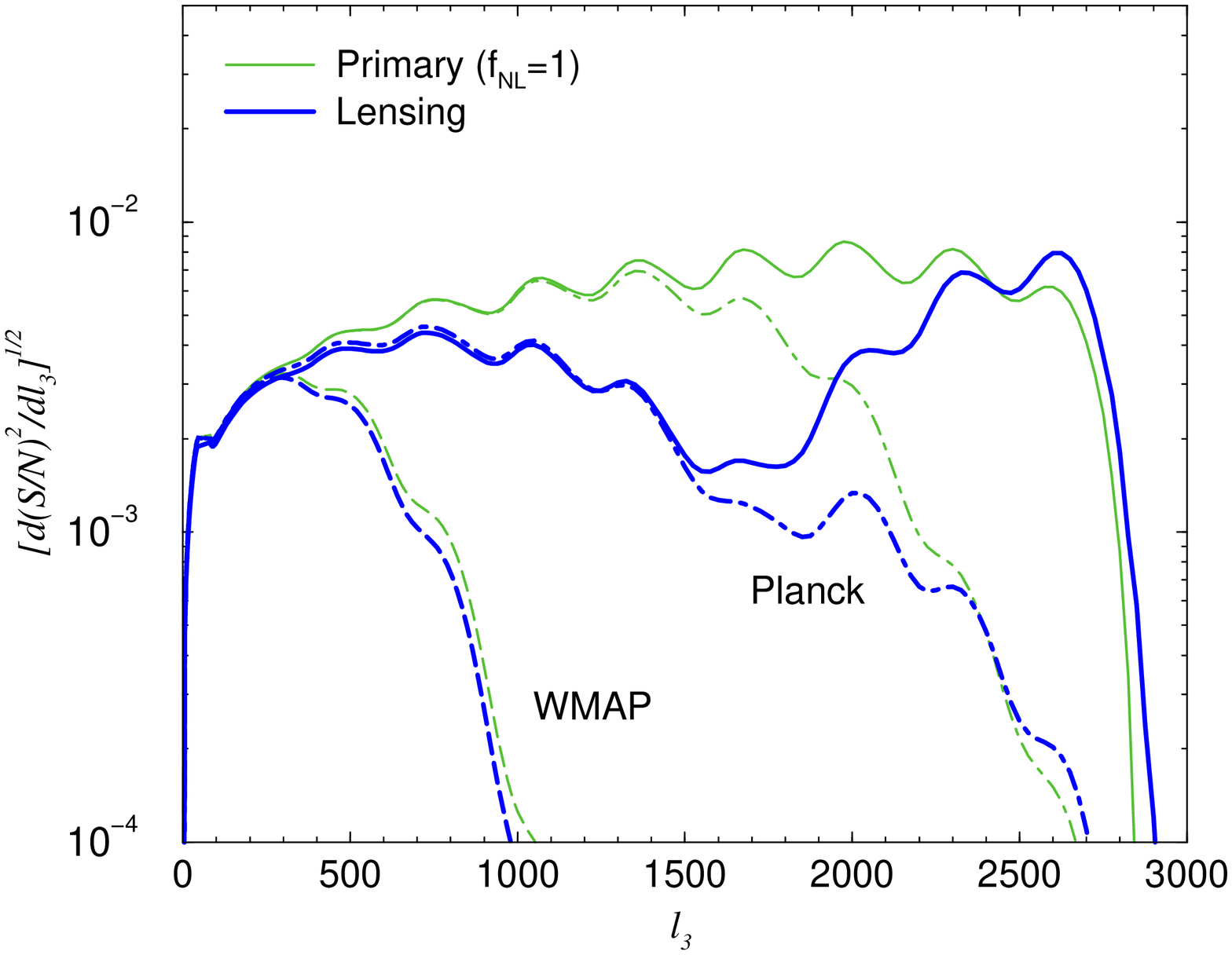}
\includegraphics[scale=0.38,angle=-90]{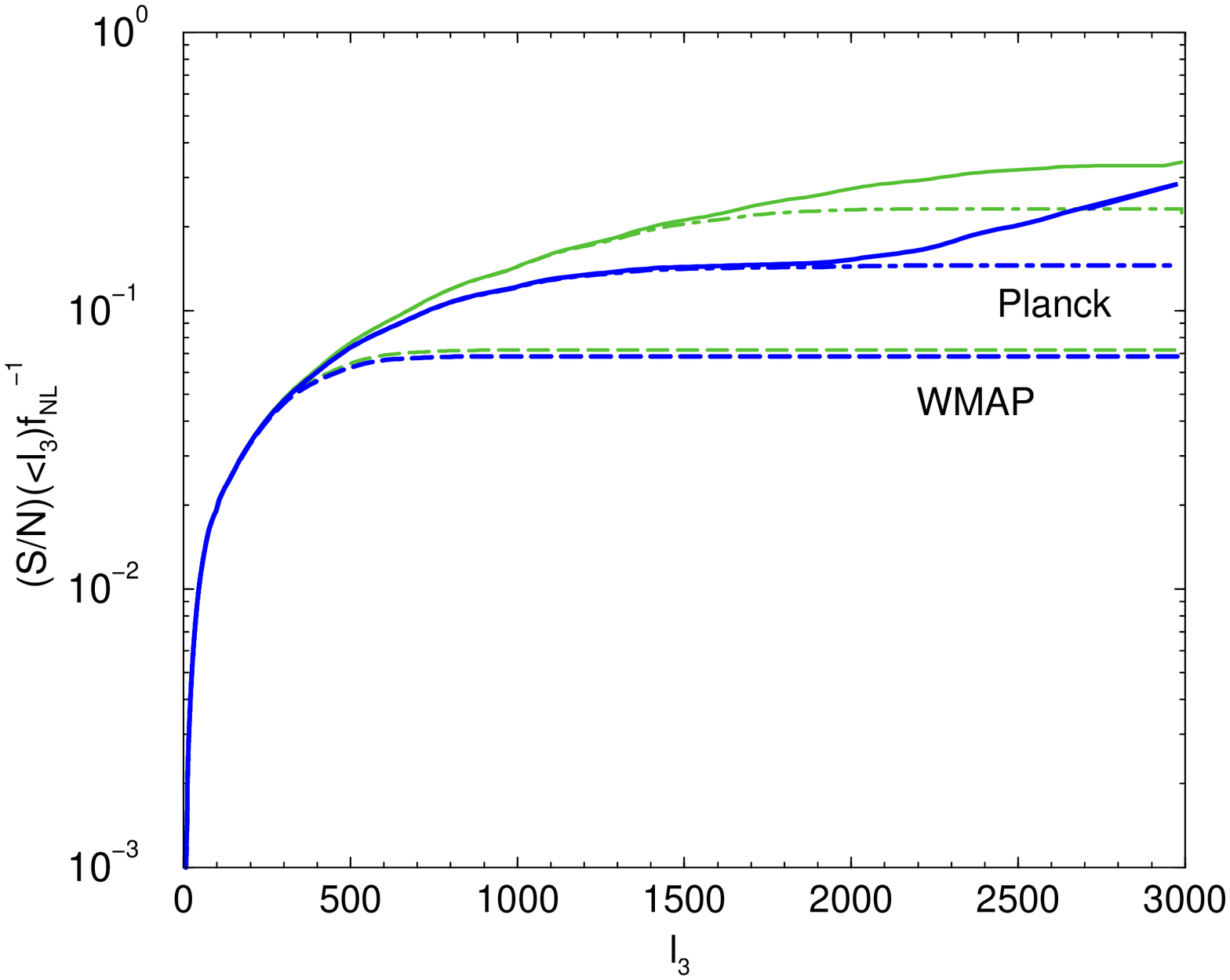}
\caption{The  signal-to-noise  ratio  for   a  detection  of  the  CMB
bispectrum with  (thick lines) and without (thin  lines) lensing.  The
long-dashed  lines show  the case  for WMAP  and dot-dashed  lines for
Planck. The left  panel shows the signal-to-noise ratio  as a function
of $l_3$,  while the right panel shows  the cumulative signal-to-noise
ratio below  $l_3$ in  the x-axis. Note  the overall reduction  in the
signal-to-noise ratio  (when $l_3 \sim  1500$) in the case  of lensing
relative to the case where lensing is ignored.}
\label{snr}
\end{figure*}

Noting that the Wigner-3$j$ symbol obeys the identity
\begin{equation}
\sum_{m_1 m_2} \wthrj{l_1}{l_2}{l_3}{m_1}{m_2}{m_3}  \wthrj{l_1}{l_2}{\lthrdp}{m_1}{m_2}{\mthrdp} = \frac{ \delta_{l_3 \lthrdp}\delta_{m_3 \mthrdp}}{(2l_3+1)},
\end{equation}
we can re-write the lensed bispectrum as 
\begin{eqnarray}
\tilde{B}_{l_1 l_2 l_3}^{\cmb} &=& {B}_{l_1 l_2 l_3}^{\cmb} + \frac{1}{2} B_{l_1 l_2 l_3}^{\cmb} \sum_{\lthrp} C_{\lthrp}^{\phi} S_1 + 2 \; \text{Perm.} \nn
&& +  \sum_{\ltwodp \lthrdp} B_{l_1  \ltwodp \lthrdp}^{\cmb} \sum_{\ltwop} C_{\ltwop}^{\phi} S_2  + 2 \; \text{Perm.} 
\label{eq:lensedbis}
\end{eqnarray}
where
\begin{eqnarray}
&& S_1 = \frac{1}{(2l_3+1)} \sum_{m_3} \sum_{\mthrpr} J_{l_3 \lthrp l_3 \lthrp}^{m_3 \mthrpr m_3 \mthrpr} \; , \; \; \text{and} \nn
&& S_2 =  \sum_{\mtwopr \mtwodp \mthrdp}  \sum_{m_1 m_2 m_3} \wthrj{l_1}{l_2}{l_3}{m_1}{m_2}{m_3} \wthrj{l_1}{\ltwodp}{\lthrdp}{m_1}{\mtwodp}{\mthrdp} \nn
&& \hspace{35mm} \times I_{l_2 \ltwop \ltwodp}^{m_2 \mtwopr \mtwodp} I_{l_3 \ltwop \lthrdp}^{m_3 \mtwopr \mthrdp} 
\end{eqnarray}
The sum of the geometric term $J_{l_3 \lthrp l_3 \lthrp}^{m_3 \mthrpr m_3 \mthrpr}$ over $\mthrpr$ yields \cite{Hu00}
\begin{equation}
S_1 = -\frac{1}{(2l_3+1)} \sum_{m_3} \frac{1}{2} l_3(l_3+1)  \lthrp (\lthrp+1) \frac{2\lthrp+1}{4\pi} \, .
\end{equation}
Hence, with summation over $m_3$ leading to a factor $(2l_3+1)$, the expression for $S_1$ simplifies to
\begin{equation}
S_1 = -\frac{1}{2} l_3(l_3+1)   \lthrp (\lthrp+1) \frac{2\lthrp+1}{4\pi}  \, .
\end{equation}

\begin{figure}[!ht]
\includegraphics[scale=0.55,angle=0]{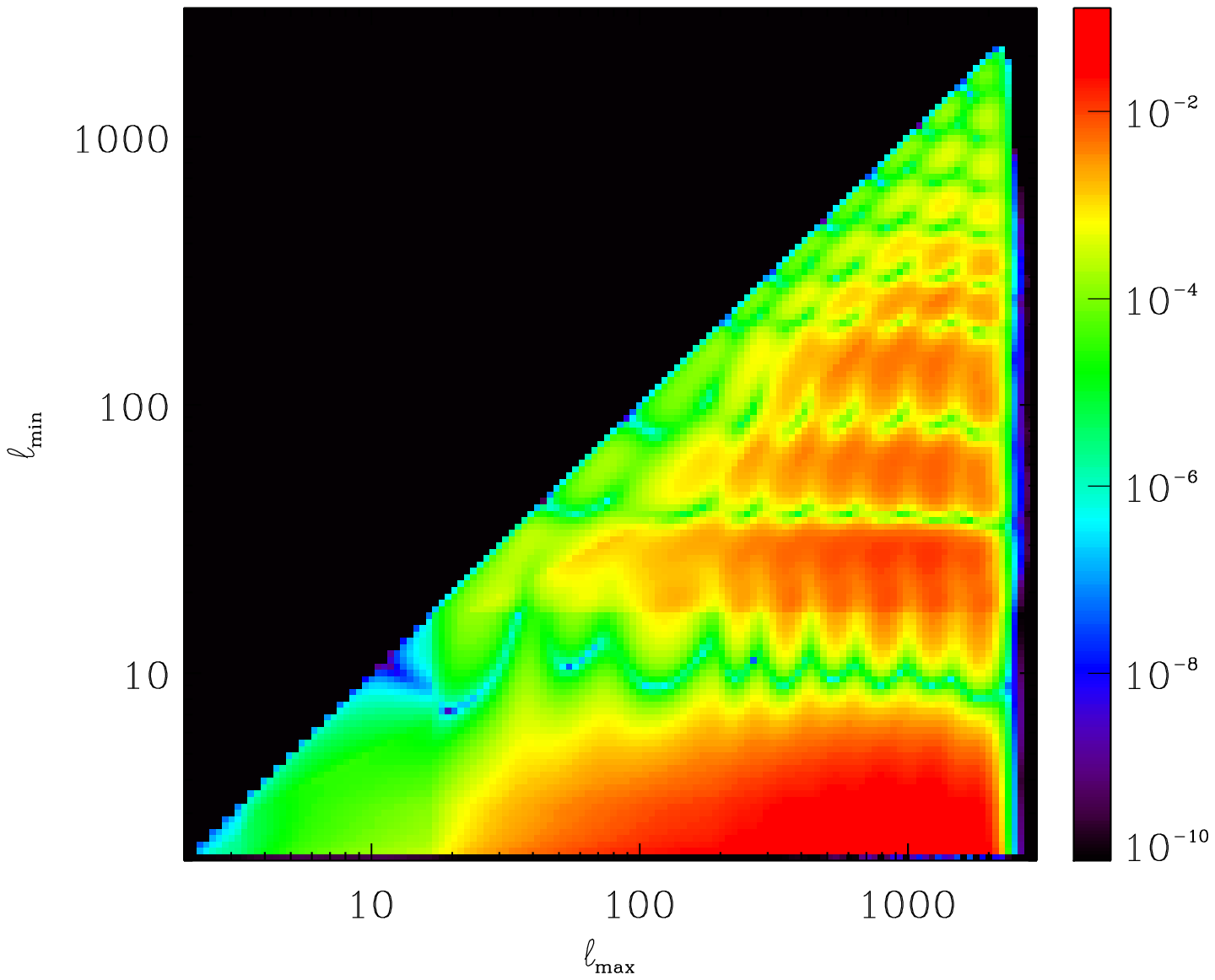}
\includegraphics[scale=0.55,angle=0]{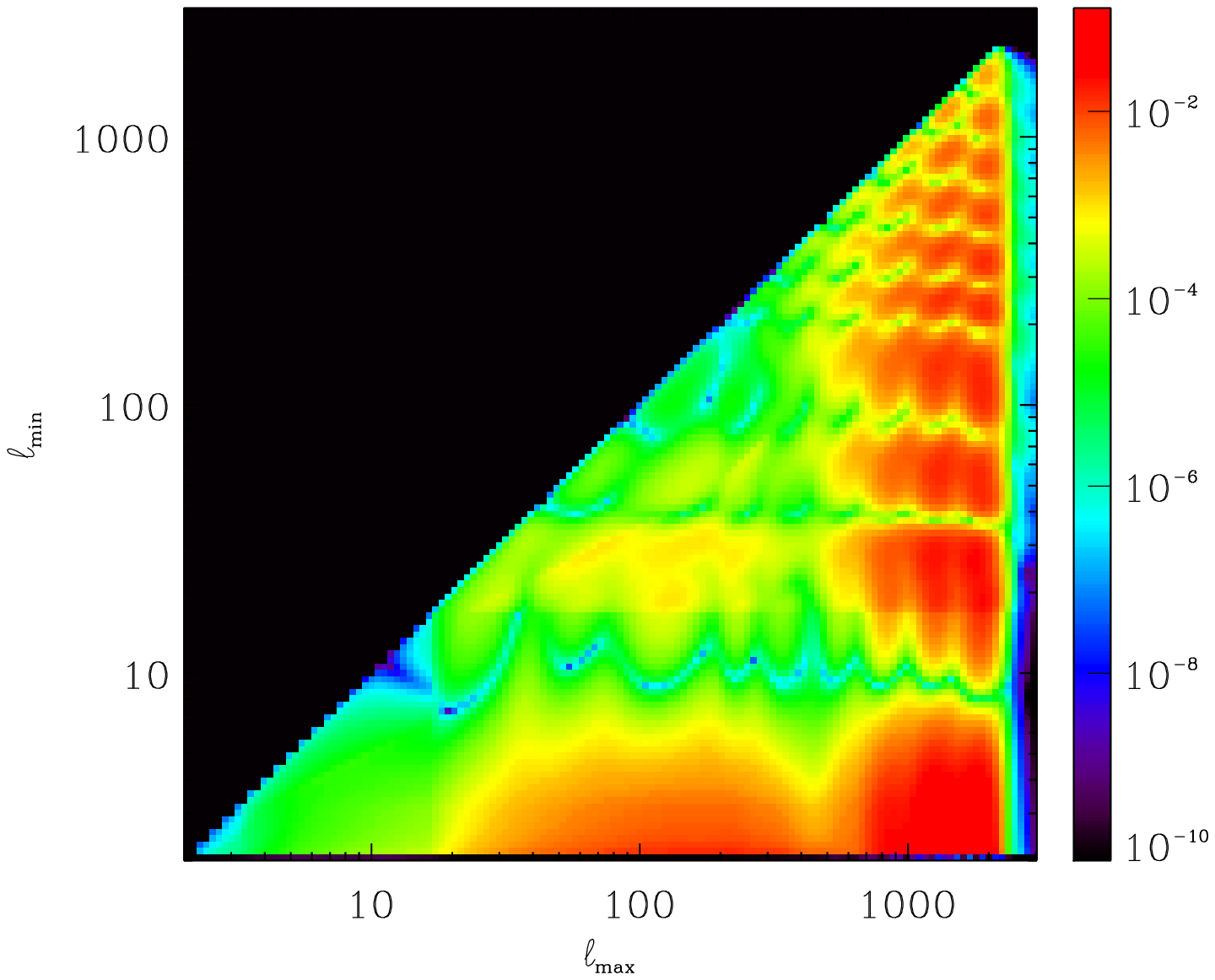}
\caption{Contour  plots  of  $d(S/N)^2/d\log l_{\rm  max}d\log  l_{\rm
min}$  (equation~\ref{eqn:F})  as  a  function of  $l_{\rm  max}$  and
$l_{\rm min}$  for the primary  bispectrum (top panel) and  the lensed
primary bispectrum (bottom panel). We take $f_{\rm NL}=1$.}
\label{Fcont}
\end{figure}

\begin{figure}[!ht]
\includegraphics[scale=0.55,angle=0]{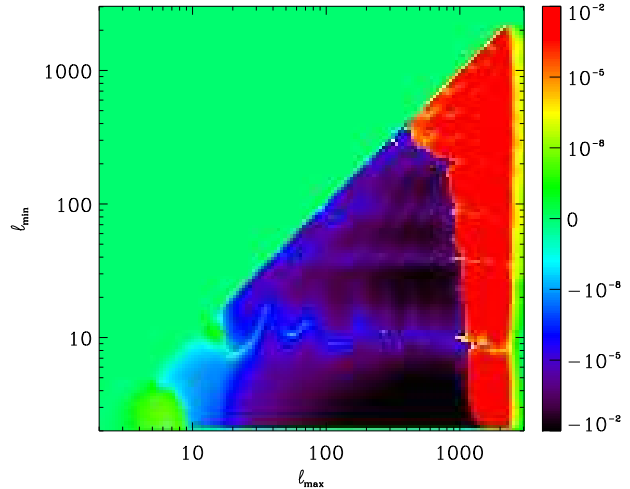}
\caption{Contour    plot    of    the    difference    $d[(S/N)^2_{\rm
lensed}-(S/N)^2_{\rm unlensed})/d\log l_{\rm max}d\log l_{\rm min}$ as
a function of $l_{\rm max}$  and $l_{\rm min}$ (same as the difference
between bottom and top panels of Figure~4).}
\label{Fdiff}
\end{figure}

\begin{figure}[!t]
\includegraphics[scale=0.38,angle=-90]{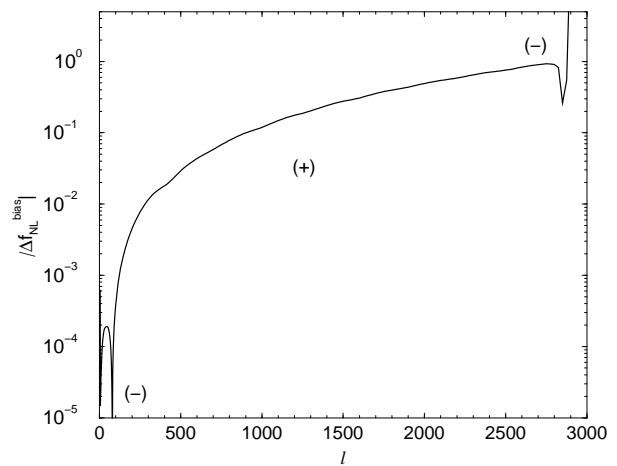}
\caption{The bias in  the primordial non-Gaussianity parameter $f_{\rm
NL}$ when lensing is ignored  in the estimation.(+)/(-) shows the sign
of $\Delta f_{\rm NL}/f_{\rm NL}$ between the true and measured values
(see text for details).}
\label{fbias}
\end{figure}

In order to evaluate $S_2$, we first re-express $I_{l \lp \ldp}^{m \mpr \mdp}$ as
\begin{equation}
I_{l \lp \ldp}^{m \mpr \mdp} = f_{l \lp \ldp} \wthrj{l}{\lp}{\ldp}{m}{\mpr}{\mdp} \, ,
\end{equation}  
where \cite{Hu00}
\begin{equation}
f_{l \lp \ldp} = \frac{1}{2} \bigg[ \lp(\lp+1) + \ldp(\ldp+1) - l(l+1)\bigg] \wthrj{l}{\lp}{\ldp}{0}{0}{0} \,.
\end{equation} 
Then the expression for $S_2$ can be re-written as
\begin{eqnarray}
S_2 &=& f_{l_2 \ltwop \ltwodp} \; f_{l_3 \ltwop \lthrdp} \sum_{\mtwopr \mtwodp \mthrdp}  \sum_{m_1 m_2 m_3} \wthrj{l_1}{l_2}{l_3}{m_1}{m_2}{m_3}  \times \; \; \nn
& & \wthrj{l_1}{\ltwodp}{\lthrdp}{m_1}{\mtwodp}{\mthrdp} \wthrj{l_2}{\ltwop}{\ltwodp}{m_2}{\mtwopr}{\mtwodp} \wthrj{l_3}{\ltwop}{\lthrdp}{m_3}{\mtwopr}{\mthrdp} \nn
&=& f_{l_2 \ltwop \ltwodp} \; f_{l_3 \ltwop \lthrdp} (-1)^{\lp+\ltwop+\lthrp} \wsixj{l_1}{l_2}{l_3}{\ltwop}{\lthrdp}{\ltwodp} \, ,
\end{eqnarray} 
where,  in  the  last  step,  we  have
introduced  the Wigner-6$j$  symbol  \cite{Hu01}.  The  values of  the
Wigner-6$j$  symbol  can  be  computed  numerically with  a  fast  and
efficient recursive algorithm \cite{Schulten}.

Finally,  substituting   the  expressions  for  $S_1$   and  $S_2$  in
equation~(\ref{eq:lensedbis}) and including all permutations in a single expression, we can write the lensed bispectrum as
\begin{eqnarray}
&& \tilde{B}_{l_1 l_2 l_3}^{\cmb} = \nn
&& \bigg[1 - \mathcal{R}\bigg\{l_1(l_1+1)+l_2(l_2+1) +l_3(l_3+1)\bigg\}\bigg] B_{l_1 l_2 l_3}^{\cmb}\nn
&& + \sum_{lpq} C_l^{\phi \phi}\bigg[ f_{l_2 l p}f_{l_3 l q} (-1)^{n} \wsixj{l_1}{l_2}{l_3}{l}{q}{p} {B}_{l_1 p q}^{\cmb}  \nn
&& \hspace{17mm} f_{l_3 l p}f_{l_1 l q} (-1)^{n} \wsixj{l_1}{l_2}{l_3}{p}{l}{q} {B}_{p l_2 q}^{\cmb}  \nn
&& \hspace{17mm} f_{l_1 l p}f_{l_2 l q} (-1)^{n} \wsixj{l_1}{l_2}{l_3}{q}{p}{l} {B}_{p q l_3}^{\cmb}  \bigg]\nonumber
\end{eqnarray}
where
\begin{equation}
\mathcal{R} = \frac{1}{4} \sum_{l} l(l+1) \frac{2l+1}{4\pi} C_{l}^{\phi \phi} 
\end{equation}
and $n\equiv (l+p+q)$.

\section{Results \& Discussion}

We illustrate  the modification  to the equilateral  configurations of
the bispectrum in Figure~1, where we plot $l^4B^\cmb_{lll}/(2\pi)^2$
as a function of the multipole $l$. The primary CMB bispectrum assumes
$f_{\rm  NL}=1$ and  is calculated  with the  full  radiation transfer
function $g_{Tl}(k)$. The lensing description makes use of the all-sky
treatment to calculate both  $C_l^\phi$ and the lensed bispectrum. The
flat-sky  expression  gives  a  result  consistent  with  the  all-sky
expression  at  better than  5\%  at  all  multipoles if  the  all-sky
expression  for   $C_l^\phi$  is  used  in   both  calculations.   The
difference is at  the level of 10\% if the two  expression make use of
the two  separate calculations  of $C_l^\phi$, as  in the case  of the
angular power  spectrum \cite{Hu00}.  In  the case of  the equilateral
configurations  of the  bispectrum,  the lensing  effect  can be  best
described  as  a  smoothing  and   a  decrease  of  the  amplitude  of
non-Gaussianity  power   in  the  equilateral   configuration  of  the
bispectrum when $l < 1500$ and a subsequent increase in the bispectrum
amplitude at small angular scales.

In Figure~2,  we show  two squeezed configurations  ($l_1\sim l_3
 \gg l_2$)  of the bispectrum,  with the short length  fixed at
 either 10 (left  panel) or 100 (right panel),  as a function of
 the multipole  $l$ of one  side with the  third side fixed  at either
 $l+10$  or $l+100$.  Without  lensing, a  comparison of  Figure~1 and
 Figure~2 reveals a well-known result in the literature that the local
 type of  the CMB bispectrum  is dominated by  squeezed configurations
 with one small side and two large sides for the bispectrum mode shape
 in the multipole space.  With  lensing, the amplitude of the squeezed
 configurations is  significantly reduced  when two of  the sides
 have lengths $l  > 1200$ in the multipole space.   This can be again
 described  as a  smoothing  effect with  lensing  by the  intervening
 large-scale  structure; removing  or ``washing  out''  the primordial
 non-Gaussian signature in  the CMB map at angular  scales below a few
 arcminutes.   Thus, when  lensed  by the  large-scale structure,  the
 primordial  non-Gaussian  CMB  sky  appears  more  Gaussian  at
 arcminute scales when studying the non-Gaussianity at the three-point
 level.   At   same  angular   scales,  however,  the   CMB  sky
 appears  more non-Gaussian  due  to lensing  at the  four-point
 level probed by the trispectrum \cite{Hu01}.

The  removal  of  the  non-Gaussianity  is  associated  with  the
squeezed  configurations, which  dominate the  overall signal-to-noise
ratio for  the detection of  the primary bispectrum  without lensing.
Although   non-Gaussianity   is   reduced  for   the   squeezed
configurations, lensing leads  to an increase in the  amplitude of the
bispectrum for equilateral configurations where $l_1 \sim l_2 \sim l_3
> 1500$.  This  increase, however,  is insignificant in  terms of
the  overall   signal-to-noise  ratio  as  the   contribution  to  the
cumulative signal-to-noise coming  from these configurations is lower,
owing to the higher variances  associated with foregrounds and instrumental
noise at these angular scales.

To  further  quantify  this  statement,  we  plot,  in  Figure~3,  the
signal-to-noise ratio calculated as
\begin{equation}
\left(\frac{S}{N}\right)^2 = \sum_{l_1l_2l_3} \frac{\left(B^\cmb_{l_1l_2l_3}\right)^2}{6C_{l_1}^{\rm tot}C_{l_2}^{\rm tot}C_{l_3}^{\rm tot}} \, ,
\end{equation}
where the noise variance calculation involves all contributions to the
angular  power   spectrum  with  $C_l^{\rm   tot}=\tilde  C_l+C_l^{\rm
sec}+N_l$  where we  include the  lensed CMB  power  spectrum ($\tilde
C_l$), secondary  anisotropies ($C_l^{\rm sec}$),  and the noise
power  spectrum   ($N_l$)  for   both  WMAP  and   Planck.   For
secondaries, we include the SZ power spectrum calculated with the halo
model  \cite{Cooray2} and  make  use of  the  noise calculations  from
Ref.~\cite{Cooray}  for WMAP and  Planck.  For  the case  involving an
experiment limited by  the cosmic variance, we set  $N_l=0$.  In
the left panel of Figure~3, we plot $d(S/N)^2/dl_3$ as a function of
$l_3$, while in the right panel we plot the cumulative signal-to-noise
ratio out to $l_3$  in the x-axis.  In the case where lensing is not included,
 signal-to-noise ratio estimates for the bispectrum detection 
are consistent with previous calculations in the literature
\cite{Komatsu}.

With lensing, however, the  signal-to-noise ratios are changed. As can
be seen  from the  left panel of  Figure~3, there is  an overall
reduction  in the  signal-to-noise ratio  when $l_3  \sim  1500$. This
difference  comes  from  the  previously  described  decrease  in  the
amplitude  of  the  non-Gaussianity in  squeezed  configurations of the bispectrum with lensing
imposed.  To  further understand the
differences in  the signal-to-noise ratio of  the lensed primary
bispectrum, we plot, in  Figure~4, the quantity
\begin{equation}
\frac{d (S/N)^2}{d\log l_{\rm max} d\log l_{\rm min}}=l_{\rm max}l_{\rm min} \sum_{l=l_{\rm min}}^{l_{\rm max}} 
\frac{\left(B^\cmb_{l_{\rm min}ll_{\rm max}}\right)^2}{6C_{l_{\rm min}}^{\rm tot}C_{l}^{\rm tot} C_{l_{\rm max}}^{\rm tot}} \, ,
\label{eqn:F}
\end{equation}
 with  two separate  estimates  for $B^\cmb$  and  $\tilde B^\cmb$  to
estimate this  quantity without and with  lensing, respectively.  Note
that  the overall  signal-to-noise ratio  comes from  integrating this
quantity over  the variables  $l_{\rm max}$ and  $l_{\rm min}$  and we
include  the  factor  $l_{\rm  max}l_{\rm  min}$ to  account  for  the
logarithmic  scaling.  A  comparison  of the  two  panels in  figure~4
reveals an overall  decrease in the amplitude at $l  \sim 10^3$ in the
case  with  the lensed  primary  bispectrum  relative  to the  primary
bispectrum alone.

In figure~5,  we plot the difference of  the two as a  contour plot to
show   that  lensing   results   in  an   overall   decrease  in   the
signal-to-noise ratio in the squeezed configurations when $l_{\rm max}
\sim 10^3$ and $l_{\rm min} < 10^2$, while there is an increase in the
signal-to-noise  ratio when  $l_{\rm  max} \sim  3\times10^3$ for  all
values  of $l_{\rm  min}$. These  plots demonstrates  the  same trends
described  with  respect  to  figure~2  involving a  decrease  in  the
amplitude   of    the   squeezed   configurations    of   the   lensed
bispectrum. While there is a  slight increase in the lensed bispectrum
amplitude at $l_3 > 2000$, such small angular scales are not probed by
Planck. Even  in the cosmic variance limit,  unfortunately, this small
increase  is insignificant  given that  at these  same  angular scales
secondary  anisotropies  dominate the  bispectrum  noise variance.  In
terms  of  the  cumulative   signal-to-noise  ratio  values  shown  in
figure~3,  the minimum  $f_{\rm  NL}$ to  detect  the bispectrum  with
Planck and a cosmic variance  limited experiment is increased by about
30\%   to  40\%  from   $f_{\rm  NL}\sim5$   for  Planck   to  $f_{\rm
NL}\sim7$. The cosmic variance-limited detection threshold for $f_{\rm
NL}$ is increased from 3 to 5.

While  the  difference   in  cumulative  signal-to-noise  ratio  seems
insignificant for  an experiment like WMAP, weak  lensing could impact
existing  measurements of the  non-Gaussianity parameter  $f_{\rm NL}$
\cite{Komatsu5yr,Yadav}.  To understand the lensing bias introduced to
$f_{\rm NL}$, we follow  the discussion in Ref.~\cite{Serra} and
note that  the current estimators  of the non-Gaussianity parameter use
$\hat{f}_{\rm      NL}       =      \frac{\hat{S}_{\rm      prim}}{N}$
\cite{Komatsu2,Creminelli,Yadav2}, with
\begin{equation}
\hat{S}_{\rm prim} = \sum_{{\bf p} {\bf q}} B^\cmb_{l_1l_2l_3} {\cal C}^{-1}_{{\bf p} {\bf q}}\hat{B}^{\rm obs}_{l_1' l_2' l_3'}\,
\label{eqn:S}
\end{equation}
where $B^\cmb_{l_1  l_2 l_3}$ is the primary
bispectrum, and  ${\cal C}_{{\bf p}{\bf q}}$ is  the covariance matrix
for  bispectrum measurements  involving  triplets of  ${\bf p}  \equiv
(l_1l_2l_3)$ and  ${\bf q} \equiv (l_1'l_2'l_3')$.   This estimator is
the  optimal  estimator for  non-Gaussianity  measurements, but  given
the  complications associated  with  estimating the  covariance,
existing   studies  make   use  of   a  sub-optimal   estimator  which
approximates      the     covariance     with      variance     ${\cal
C}^{-1}_{\mathbf{\alpha}\mathbf{\alpha'}}\approx
\sigma^{-2}(l_1,l_2,l_3)\delta_{\mathbf{\alpha}\mathbf{\alpha'}}$,
and introduces a linear term to equation~(\ref{eqn:S}) to minimize the
variance of  $\hat{f}_{\rm NL}$ \cite{Yadav2}.   Note that $N$  is the
overall   normalization   factor   that   can   be   calculated   from
equation~(\ref{eqn:S}) by replacing $\hat{B}^{\rm obs}$ with $B^\cmb$.

While  weak  lensing modifies  the  observed bispectrum  $\hat{B}^{\rm
obs}_{l_1  l_2  l_3}=f_{\rm  NL}\tilde  B^\cmb_{l_1l_2l_3}$,  existing
measurements make the assumption that $\hat{B}^{\rm obs}_{l_1 l_2 l_3}
= f_{\rm NL} B_{l_1 l_2  l_3}^\cmb$. This results in a biased estimate
of  $\hat{f}_{\rm NL}$  from  the true  value  of the  non-Gaussianity
parameter $f^{\rm  true}_{\rm NL}$. The fractional  difference of this
bias    $\Delta    f/\hat{f}_{\rm    NL}\equiv   (f^{\rm    true}_{\rm
NL}-\hat{f}_{\rm NL})/\hat{f}_{\rm NL}$  can be calculated through the
covariance  between the  lensed  and unlensed  CMB primary  bispectrum
$\Delta  f/\hat{f}_{\rm  NL}=  1-[\sum B^\cmb_{l_1l_2l_3}  \sigma^{-2}
\tilde  B^\cmb_{l_1  l_2  l_3}/  \sum  B^\cmb_{l_1l_2l_3}  \sigma^{-2}
B^\cmb_{l_1 l_2  l_3}]$, where we  have simply written the  variance as
$\sigma^{-2}$.  We  plot $\Delta f/\hat{f}_{\rm NL}$ as  a function of
$l$ to  which non-Gaussianity parameter measurements  are performed in
figure~6. Existing  measurements with WMAP  data probe out  to $l_{\rm
max} \sim 750$ and we find that existing estimates of $f_{\rm NL}$ are
biased by $\sim$ 6\%.  For Planck,  if lensing is ignored, the bias is
at the level of 30\%.

This bias is not the same fractional difference in the signal-to-noise
ratio that one can infer from figure~3 since the fractional difference
in the signal-to-noise ratio with and without lensing involves a ratio
of  the  form  $[\sum  \tilde  B^\cmb_{l_1l_2l_3}  \sigma^{-2}  \tilde
B^\cmb_{l_1  l_2 l_3}/\sum B^\cmb_{l_1l_2l_3}  \sigma^{-2} B^\cmb_{l_1
l_2 l_3}]$.  Note that in  future an unbiased estimate of $f_{\rm NL}$
can    be    obtained    by    replacing    $B^\cmb_{l_1l_2l_3}$    in
equation~(\ref{eqn:S})    with   the    lensed    bispectrum   $\tilde
B^\cmb_{l_1l_2l_3}$  and recalculating  the  normalization factor  $N$
with  the  lensed  primary  bispectrum. Unfortunately,  while  without
lensing  the CMB  primary  bispectrum of  the  local model  factorizes
into  two separate  integrals with $b_l^L$  and $b_l^{\rm
NL}$ (described  in Section~II) ,  this factorizability is  no longer
preserved  when   lensed  and  impacts  an  easy   estimation  of  the
non-Gaussianity     parameter    with    the     existing    estimator
\cite{Komatsu2}. For  Planck and  other CMB experiments  that can
probe  down to  small angular  scales for  primordial non-Gaussianity
measurements,  it will  be necessary  to implement  an  estimator that
accounts for the lensing effect.

To summarize  the main  results of this  paper, we have  discussed the
primary CMB bispectrum generated at the
last scattering surface, but observed today after it is weak lensed by
the  intervening  large-scale  structure.  Unfortunately, as  we  have
found, weak lensing  leads to an overall decrease  in the amplitude of
non-Gaussianity with the biggest change on the squeezed configurations
of the bispectrum that dominate the overall signal-to-noise ratio when
studying the primordial  non-Gaussianity parameter.  For an experiment
such  as   the  Wilkinson  Microwave  Anisotropy   Probe  (WMAP),  the
modifications  imposed by lensing  results in  an estimate  of $f_{\rm
NL}$ of the local  model that is biased low by about  6\%.  For a high
resolution experiment such as  Planck, the lensing modification to the
bispectrum  must be  accounted  for when  attempting  to estimate  the
primordial non-Gaussianity.   The minimum detectable  value of $f_{\rm
NL}$ for a cosmic variance limited experiment is $\sim$ 5.

\begin{center}
{\bf Acknowledgments}
\end{center}

We  thank  participants  of  the recent  non-Gaussianity  workshop  at
the Perimeter Institute for helpful  discussions and Alex Amblard for help with Figure~4. 
PS thanks Tommaso Stasi. This  work   was  supported  by   NSF  CAREER AST-0645427.  We acknowledge  the use  of CMBFAST  by Uros  Seljak and
Matias Zaldarriaga \cite{Seljak}.


\begin{thebibliography}{99}

\bibitem{lensing}  See, e.g., U.~Seljak and M.~Zaldarriaga, \PRL\ {\bf                                                                                         
  82}, 2636 (1999); \PRD\ {\bf 60},
  043504 (1999); M. Zaldarriaga and
  U. Seljak, \PRD\ {\bf 59}, 123507 (1999); 
  M.~Zaldarriaga,
  Phys.\ Rev.\  D {\bf 62}, 063510 (2000)
  [arXiv:astro-ph/9910498];
M.~H.~Kesden, A.~Cooray and M.~Kamionkowski,
  Phys.\ Rev.\  D {\bf 66}, 083007 (2002)
  [arXiv:astro-ph/0208325];
C.~Vale, A.~Amblard and M.~J.~White,
  New Astron.\  {\bf 10}, 1 (2004)
  [arXiv:astro-ph/0402004];
A.~Amblard, C.~Vale and M.~J.~White,
  New Astron.\  {\bf 9}, 687 (2004)
  [arXiv:astro-ph/0403075];
  S.~Das and P.~Bode,
  arXiv:0711.3793 [astro-ph].

\bibitem{Lewis}
  For a recent review see, A.~Lewis and A.~Challinor,
  Phys.\ Rept.\  {\bf 429}, 1 (2006)
  [arXiv:astro-ph/0601594].

\bibitem{Hu00}
  W. Hu,  Phys. Rev. D\ {\bf 62}  043007 (2000)
  [arXiv:astro-ph/0001303]


\bibitem{HuOka02}
     W.~Hu and T.~Okamoto, \ApJ\ {\bf 574}, 566 (2002)
     [arXiv:astro-ph/0111606];
M.~Kesden, A.~Cooray, and M.~Kamionkowski,
        \PRD\ {\bf 67}, 123507 (2003) [arXiv:astro-ph/0302536];
C.~M.~Hirata and U.~Seljak,
Phys.\ Rev.\ D {\bf 68}, 083002 (2003)
[arXiv:astro-ph/0306354].

\bibitem{KamKosSte97} M.~Kamionkowski, A.~Kosowsky, and
     A.~Stebbins, \PRL\ {\bf 78}, 2058 (1997) [arXiv:astro-ph/9609132];
        U.~Seljak and M.~Zaldarriaga, \PRL\ {\bf 78}, 2054
     (1997) [arXiv:astro-ph/9609169].

\bibitem{KesCooKam02} M.~Kesden, A.~Cooray, and M.~Kamionkowski,
     \PRL\ {\bf 89}, 011304 (2002) [arXiv:astro-ph/0202434];
     L.~Knox and Y.-S.~Song, \PRL\ {\bf 89}, 011303
     [arXiv:astro-ph/0202286]; U. Seljak and C. Hirata,
     Phys. Rev. D {\bf 69}, 043005 (2004) [arXiv:astro-ph/0310163].


\bibitem{Zaldarriaga:1998ar}
  M.~Zaldarriaga and U.~Seljak,
  Phys.\ Rev.\ D {\bf 58}, 023003 (1998)
  [arXiv:astro-ph/9803150].


\bibitem{Smith}
  K.~M.~Smith, O.~Zahn and O.~Dore,
  Phys.\ Rev.\  D {\bf 76}, 043510 (2007)
  [arXiv:0705.3980 [astro-ph]].


\bibitem{Hirata}
  C.~M.~Hirata, S.~Ho, N.~Padmanabhan, U.~Seljak and N.~Bahcall,
  arXiv:0801.0644 [astro-ph].

\bibitem{Komatsu}
  E.~Komatsu and D.~N.~Spergel,
  Phys.\ Rev.\ D {\bf 63}, 063002 (2001)
  [arXiv:astro-ph/0005036].


\bibitem{Lig}
  M.~Liguori, F.~K.~Hansen, E.~Komatsu, S.~Matarrese and A.~Riotto,
  Phys.\ Rev.\ D {\bf 73}, 043505 (2006)
  [arXiv:astro-ph/0509098].


\bibitem{Komatsu4}
  E.~Komatsu {\it et al.}  [WMAP Collaboration],
  Astrophys.\ J.\ Suppl.\  {\bf 148}, 119 (2003)
  [arXiv:astro-ph/0302223].


\bibitem{Komatsu5yr}
 E.~Komatsu {\it et al.}  [WMAP Collaboration],
  arXiv:0803.0547 [astro-ph].


\bibitem{Yadav}
  A.~P.~S.~Yadav and B.~D.~Wandelt,
  [arXiv:0712.1148].


\bibitem{Salopek}
D.~S.~Salopek and J.~R.~Bond,
Phys.\ Rev.\ D {\bf 42}, 3936 (1990); ibid. {\bf 43}, 1005 (1991)

\bibitem{Falk}
T.~Falk, R.~Rangarajan and M.~Srednicki,
Astrophys.\ J.\  {\bf 403}, L1 (1993)

\bibitem{Gangui}
A.~Gangui, F.~Lucchin, S.~Matarrese and S.~Mollerach,
Astrophys.\ J.\  {\bf 430}, 447 (1994)

\bibitem{Pyne}
T.~Pyne and S.~M.~Carroll,
Phys.\ Rev.\ D {\bf 53}, 2920 (1996)


\bibitem{Maldacena}
  J.~M.~Maldacena,
  JHEP {\bf 0305}, 013 (2003).


\bibitem{Acquaviva}
V.~Acquaviva, N.~Bartolo, S.~Matarrese and A.~Riotto,
Nucl. Phys. {\bf B667}, 119 (2003), 
[arXiv:astroph-/0209156]

\bibitem{Bartolo}
N.~Bartolo, S.~Matarrese and A.~Riotto,
  Phys.\ Rev.\ Lett.\  {\bf 93}, 231301 (2004)
  [arXiv:astro-ph/0407505].

 \bibitem{Bartolo2}
N.~Bartolo, E.~Komatsu, S.~Matarrese and A.~Riotto,
Phys.~Rept. {\bf 402} (2004) 103-266
                                             

\bibitem{Buchbinder}
  E.~I.~Buchbinder, J.~Khoury and B.~A.~Ovrut,
  arXiv:0710.5172 [hep-th].

\bibitem{Lehners}
  J.~L.~Lehners and P.~J.~Steinhardt,
  arXiv:0712.3779 [hep-th].

\bibitem{Goldberg}
  D.~M.~Goldberg and D.~N.~Spergel,
  Phys.\ Rev.\  D {\bf 59}, 103002 (1999)
  [arXiv:astro-ph/9811251].

\bibitem{Cooray}
  A.~R.~Cooray and W.~Hu,
  Astrophys.\ J.\  {\bf 534}, 533 (2000)
  [arXiv:astro-ph/9910397].


\bibitem{Smith2}
  K.~M.~Smith and M.~Zaldarriaga,
  arXiv:astro-ph/0612571.

\bibitem{Serra}
  P.~Serra and A.~Cooray,
  arXiv:0801.3276 [astro-ph].


\bibitem{BabichPier}
  D.~Babich and E.~Pierpaoli,
  arXiv:0803.1161 [astro-ph].

\bibitem{Komatsu2}
  E.~Komatsu, D.~N.~Spergel and B.~D.~Wandelt,
  Astrophys.\ J.\  {\bf 634}, 14 (2005)
  [arXiv:astro-ph/0305189].

\bibitem{Seljak}
  U.~Seljak and M.~Zaldarriaga,
  Astrophys.\ J.\  {\bf 469}, 437 (1996)
  [arXiv:astro-ph/9603033].


\bibitem{cosmicshearrefs} R. D. Blandford et al.,
     Mon.\ Not.\ Roy.\ Astron.\ Soc.\ {\bf 251}, 600 (1991);
     J. Miralda-Escud\'e, Astrophys.\ J.\ {\bf 380}, 1 (1991);
     N.~Kaiser, Astrophys.\ J.\  {\bf 388}, 277 (1992);
     M. Bartelmann and P. Schneider, Astron.\ Astrophys.\ {\bf                                                                                                                              
     259}, 413 (1992).
For reviews, see,                                                                                                                                                                          
M.~Bartelmann and P.~Schneider, Phys. Rept. {\bf 340}, 291 (2001);                                                                                                                         
P.~Schneider  Gravitational Lensing: Strong, Weak \& Micro,                                                                                                                                
Lecture Notes of the 33rd Saas-Fee Advanced Course, (Berlin:                                                                                                                               
Springer-Verlag).                 

\bibitem{CooHutime}
  W.~Hu and A.~Cooray,
  Phys.\ Rev.\  D {\bf 63}, 023504 (2001)
  [arXiv:astro-ph/0008001].

\bibitem{Cooray3}
  A.~R.~Cooray,
  New Astron.\  {\bf 9}, 173 (2004)
  [arXiv:astro-ph/0309301];
   A.~Challinor and A.~Lewis,
  Phys.\ Rev.\  D {\bf 71}, 103010 (2005)
  [arXiv:astro-ph/0502425].

\bibitem{Hu01}
W.~Hu, \PRD\ {\bf 64}, 083005 (2001).

\bibitem{Schulten}
K.~Schulten and R.~Gordon, J. Math. Phys.\ {\bf 16}, 1971 (1975).

\bibitem{Cooray2}
  A.~Cooray and R.~K.~Sheth,
  Phys.\ Rept.\  {\bf 372}, 1 (2002)
  [arXiv:astro-ph/0206508];
  A.~Cooray,
  Phys.\ Rev.\  D {\bf 62}, 103506 (2000)
  [arXiv:astro-ph/0005287].

\bibitem{Creminelli}
  P.~Creminelli, A.~Nicolis, L~Senatore, M.~Tegmark and M.~Zaldarriaga, 
Journal of Cosmology and Astro-Particle Physics {\bf 5}, 4 (2006),
[arXiv:astro-ph/0509029]
 
\bibitem{Yadav2}
  A.~P.~S.~Yadav, E.~Komatsu, B.~D.~Wandelt, M.~Liguori, F.~K.~Hansen and S.~Matarrese,
  arXiv:0711.4933 [astro-ph].


\end{thebibliography}
\end{document}